\documentclass[twocolumn,showpacs,preprintnumbers,amsmath,amssymb]{revtex4}
\usepackage{graphicx}
\usepackage{dcolumn}
\usepackage{bm}

\begin{document}

\draft
\title{Electric-field dependent spin diffusion
and spin injection into semiconductors}

\author{Z. G. Yu and M. E. Flatt\'e}

\address{Department of Physics and Astronomy, University of Iowa,
Iowa City, Iowa 52242}
\date{\today}

\begin{abstract}
We derive a drift-diffusion equation for spin polarization
in semiconductors by consistently taking into account
electric-field effects and nondegenerate electron statistics.
We identify a high-field diffusive regime which has no analogue
in metals. In this regime there are two distinct
spin diffusion lengths. Furthermore, spin injection
from a ferromagnetic
metal  into a semiconductor is enhanced by several orders
of magnitude and spins can be transported over distances
much greater than the low-field spin diffusion length.
\end{abstract}
\phantom{.}

\maketitle

Semiconductor devices based on the control and manipulation
of electron spin (semiconductor spintronics)
 have recently attracted considerable attention \cite{wolf}.
Spin transport and injection properties of semiconductors
and heterostructures strongly constrain the design of new
spintronic devices.
In theoretical
studies of spin transport and injection in semiconductors \cite{schmidt, rashba, smith}
the spin polarization is usually assumed to obey the same diffusion
equation as in metals \cite{son},
\begin{equation}
\nabla^2 (\mu_{\uparrow}-\mu_{\downarrow})
-(\mu_{\uparrow}-\mu_{\downarrow})/L^2=0,
\end{equation}
where $\mu_{\uparrow(\downarrow)}$ is the electrochemical potential
of up-spin (down-spin) electrons.
In this diffusion equation, the
electric field does not play any role, and spin polarization
decays away on a length scale of $L$ from an injection point.
This is reasonable
for metals because
the electric field {\bf E} is essentially screened.
For semiconductor spintronic devices, however,
the semiconductor often
is lightly doped and nondegenerate, and moderate electric field
can dominate the carrier motion. Equation (1) corresponds to
neglecting drift
in the more general
 drift-diffusion equation for the spin polarization,
\begin{equation}
\nabla^2(n_{\uparrow}-n_{\downarrow})+\frac{e{\bf E}}{k_B T}
\cdot{\bf\nabla}(n_{\uparrow}-n_{\downarrow})
-\frac{(n_{\uparrow}-n_{\downarrow})}{(L^{(s)})^2}=0,
\end{equation}
where $n_{\uparrow}-n_{\downarrow}$ is the difference between
up-spin and down-spin electron densities and $L^{(s)}$ is the intrinsic
spin diffusion length.

If Eq. (1) holds,
spin injection
from a ferromagnetic metal to a semiconductor without  a
spin-selective
 interfacial barrier
 is virtually impossible
due to the ``conductivity
mismatch'', or more precisely,
a mismatch between effective
resistances in the metal ($L^{(f)}/\sigma_f$) and in the semiconductor
($L^{(s)}/\sigma_s$) \cite{schmidt,rashba,smith}.
Here $L^{(f)}$ and $L^{(s)}$  are the spin diffusion lengths
for the ferromagnetic metal and the semiconductor, and $\sigma_f$ and
$\sigma_s$ are
conductivities for the two materials.
Even for spin injection from ferromagnetic semiconductors,
$L^{(f)}/\sigma_f \gg L^{(s)}/\sigma_s$,
and the spin polarization is
much less than 99\%, so the large spin injection percentages
achieved from ZnMnSe \cite{molenkamp,jonker1}
and GaMnAs \cite{ohno} are difficult to understand via Eq. (1).

Here we clarify  the central role of the electric field
on spin transport in semiconductors. We
obtain the drift-diffusion equation (2) for the spin polarization
in a semiconductor. Equation (2) consistently takes into account
electric-field effects and nondegenerate electron statistics.
We identify a high-field diffusive regime which has no analogue
in metals. This regime occurs for field as small as 1 V/cm at low temperatures.
Two distinct
spin diffusion lengths now characterize
spin motion, i.e., up-stream ($L_u$) and down-stream ($L_d$)
spin diffusion lengths, which
can differ in orders of magnitude with realistic fields:
$E\ge 2.5$ V/cm at $T=3$ K and $E\ge 250$ V/cm at $T=300$ K.
These two length scales play distinctive but both favorable roles
in spin injection from a ferromagnetic metal to a
semiconductor. We find that
the effective semiconductor resistance determining the
injection efficiency
is $L_u/\sigma_s$
rather than $L^{(s)}/\sigma_s$, which may be comparable to
$L^{(f)}/\sigma_f$ given that $L_u$ can be shorter than $L^{(s)}$
by several orders of magnitude
in the high-field regime. Moreover, the decay length
scale for the spin polarization injected into the semiconductor is
$L_d$, which would be much longer than $L^{(s)}$ in the presence of
a strong field.
Our results suggest a simple and practical approach to increase
spin injection and spin coherence in semiconductors,
namely, increasing the electric field, or equivalently,
increasing the total injection current in semiconductors.
Our results are consistent with the
significant current dependence observed for
spin injection from Fe to GaAs \cite{jonker2}.
We further note that strong fields also substantially enhance
spin injection in structures with an interfacial
barrier.

The semiconductor we consider here is lightly or moderately
$n$-doped ($p$-doped semiconductors can be analyzed similarly),
which is typical in spintronic devices.   We assume that there
is no space charge and the material
is homogeneous.
The current for up-spin and down-spin can be written as
$
{\bf j}_{\uparrow(\downarrow)}=\sigma_{\uparrow(\downarrow)}{\bf E} +e D
{\bf \nabla} n_{\uparrow(\downarrow)},
$
which consists of the drift current and the diffusion one.
Here $D$ is the electron diffusion constant,
$\sigma_{\uparrow(\downarrow)}$ the up-spin (down-spin)
conductivity, and $n_{\uparrow(\downarrow)}$ the up-spin
(down-spin)
electron density.
The spin-dependent conductivity is proportional to the electron density
for individual spins,
$
\sigma_{\uparrow(\downarrow)}= n_{\uparrow(\downarrow)} e\nu_e,
$
where the mobility $\nu_e$ is assumed to be independent of field and
density.
The rate at which spin-up (spin-down) electrons scatter to spin-down
(spin-up)
electrons is denoted by $1/\tau_{\uparrow\downarrow}$
($1/\tau_{\downarrow\uparrow}$).
In steady state, the equations of continuity for individual spins read
\begin{eqnarray*}
{\bf \nabla}\cdot{\bf j}_{\uparrow}&=& {\bf \nabla}\sigma_{\uparrow}\cdot{\bf E}
+\sigma_{\uparrow}{\bf \nabla}\cdot{\bf E}+e D\nabla^2n_{\uparrow}
=\Bigl(\frac{n_{\uparrow}}{\tau_{\uparrow\downarrow}}-\frac{n_{\downarrow}}
{\tau_{\downarrow\uparrow}}\Bigr)e,\\
{\bf \nabla}\cdot{\bf j}_{\downarrow}&=& {\bf \nabla}\sigma_{\downarrow}\cdot{\bf E}
+\sigma_{\downarrow}{\bf \nabla}\cdot{\bf E}+e D\nabla^2n_{\downarrow}
=\Bigl(\frac{n_{\downarrow}}{\tau_{\downarrow\uparrow}}-\frac{n_{\uparrow}}
{\tau_{\uparrow\downarrow}}\Bigr)e.
\end{eqnarray*}
In nondegenerate semiconductors,
$\tau^{-1}_{\uparrow\downarrow}=\tau^{-1}_{\downarrow\uparrow}
\equiv \tau^{-1}/2$.

\begin{figure}
\vspace{5pt}
\includegraphics[width=7cm]{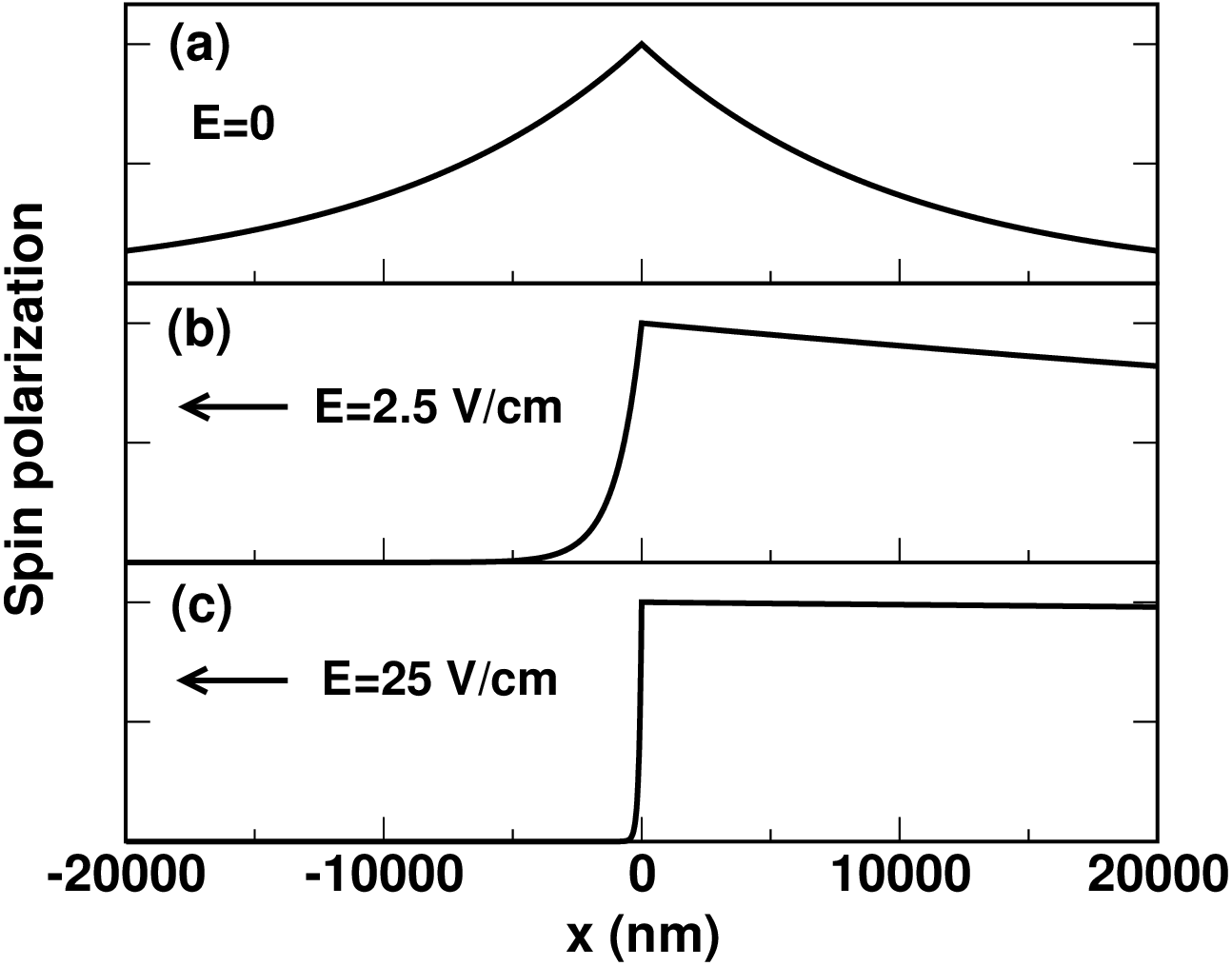}
\caption{Distribution of electron 
spin polarization as a function of position
for a spin imbalance injected at $x=0$. Panel (a), (b), and (c) are for
$|eE|/k_B T=0$, 0.001, and 0.01 nm$^{-1}$,  which correspond to
$E=0$, 2.5 V/cm, and 25 V/cm, respectively, at $T=3$ K.
The intrinsic spin diffusion length is $L^{(s)}=10^4$ nm.}
\vspace{-10pt}
\end{figure}

For a homogeneous
semiconductor without space-charge, local variation of electron
density $\Delta n$ should be balanced by a local change
of hole concentration $\Delta p$. In doped semiconductors,
spin polarization can be created without changing
electrons or hole densities
$\Delta n=\Delta p=0$ \cite{neutrality}, and therefore,
\begin{equation}
\Delta n_{\uparrow}+ \Delta n_{\downarrow}=0.
\end{equation}
  Here $\Delta n_{\uparrow(\downarrow)}=n_{\uparrow(\downarrow)}
-n_0/2$, and $n_0$ is the total electron density
in equilibrium.
From Poisson's equation,
$
{\bf \nabla}\cdot{\bf E} = -(\Delta n_{\uparrow}
+\Delta n_{\downarrow})e/\epsilon=0.
$
By using the Einstein's relation,
$D=k_B T \nu_e/e$, where $k_B$ is the Boltzmann constant and $T$ is temperature,
we obtain the differential equation (2) for $n_{\uparrow}-n_{\downarrow}$,
the measure of the spin polarization in semiconductors,
with $L^{(s)}=\sqrt{D \tau}$.

Equation (2), together with the local charge
neutrality constraint Eq. (3),
dramatically alters the spin transport behavior in semiconductors
from that expected from Eq. (1).
The general form of solution to Eq. (2) (restricting variation to the
$x$-direction)
is
$$
n_{\uparrow}-n_{\downarrow}=A \exp(-x/L_1) +B \exp (- x/L_2),
$$
where $\lambda_1=1/L_1$ and $\lambda_2 =1/L_2$ are the roots of the
quadratic equation,
$\lambda^2-\lambda eE/k_BT -1/(L^{(s)})^2=0.$
One of the roots of the above equation
must be positive and the other negative.
The choice of roots is constrained by the boundary conditions
at $\pm \infty$.
To understand the physical consequence of the electric field on the spin
transport,
we suppose that a continuous spin imbalance is injected at $x=0$, and the electric
field is along the
$-x$ direction. The spin polarization will gradually
decay in size as the distance
from the point of injection increases.
In Fig. 1, we plot the spin polarization as a function of position for
different fields. In the absence of the field, as shown in Fig. 1(a),
the spin polarization decays symmetrically along $-x$ and $+x$ with
a single length scale, $L^{(s)}$. When an electric field is applied,
the decay of the spin polarization becomes spatially asymmetric.
For spin diffusion opposite to the field direction (down-stream for
electrons),
the decay length of the spin polarization is longer than $L^{(s)}$.
For spin diffusion along the field direction (up-stream for electrons),
the decay length is shorter than $L^{(s)}$.
As we change the strength of the field, the spatial distribution
of the spin polarization can change dramatically.

\begin{figure}
\vspace{5pt}\includegraphics[width=7cm]{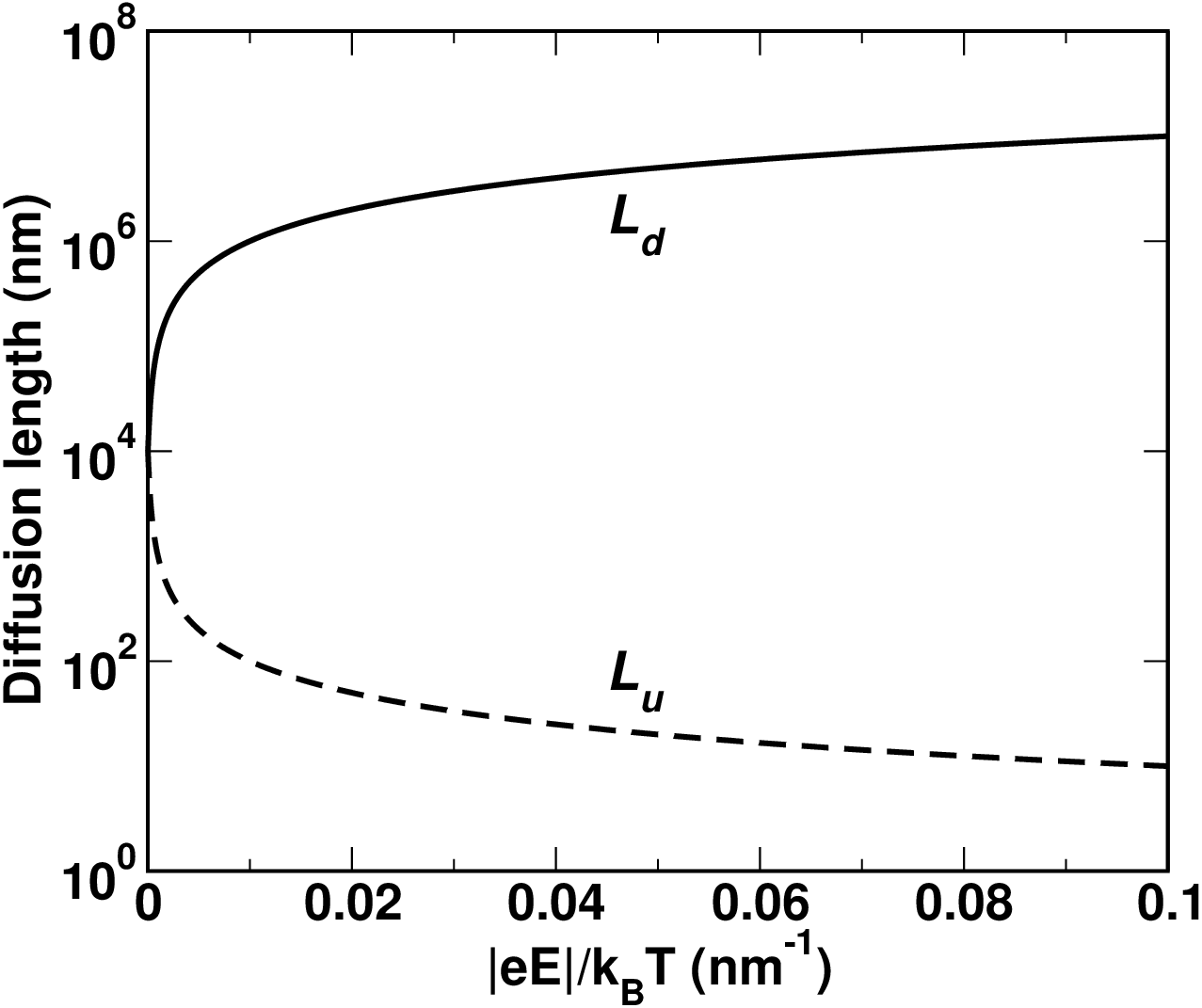}
\caption{Up-stream (dashed line) and down-stream (solid line)
diffusion lengths as a function
of electric field. The intrinsic spin diffusion length is $L^{(s)}=10^4$ nm.}
\vspace{-10pt}
\end{figure}

We define two quantities $L_d$, $L_u$,
\begin{equation}
L_d=\biggl[-\frac{|e E|}{2k_B T}+
\sqrt{\Bigl(\frac{e E}{2 k_B T}\Bigr)^2+\frac{1}{(L^{(s)})^2}}~\biggr]^{-1},
\end{equation}
\begin{equation}
L_u=\biggl[\frac{|e E|}{2k_B T}+
\sqrt{\Bigl(\frac{e E}{2 k_B T}\Bigr)^2+\frac{1}{(L^{(s)})^2}}~\biggr]^{-1}.
\end{equation}
The distribution of the spin polarization in Fig. 1 is then described
by $n_{\uparrow}-n_{\downarrow} \sim \exp (-x/L_d)$ for $x>0$,
and
$n_{\uparrow}-n_{\downarrow} \sim \exp (x/L_u)$ for $x<0$.
Thus $L_d$ ($L_d>L^{(s)}$) and $L_u$ ($L_u < L^{(s)}$)
are the down-stream and up-stream spin diffusion lengths,
respectively.

Figure 2 shows $L_d$ and $L_u$ as
a function of the electric field.
In the absence of the field,
the down-stream and up-stream lengths are equal to the intrinsic
diffusion length $L^{(s)}$. With increasing field the down-stream
diffusion length $L_d$ increases, whereas the up-stream diffusion length
$L_u$ decreases.
A high-field regime for spin transport in semiconductors
can be defined by $E > E_c$, where $eE_c/k_BT =1/L^{(s)}$.
In this regime, $L_u$ and $L_d$ deviate from $L^{(s)}$
considerably and the spin diffusion behavior
is qualitatively different from that in low fields.
We emphasize that since $L^{(s)}$ is large in semiconductors, this
regime
is not beyond realistic fields where most spintronic devices operate.
For a typical spin diffusion length, $L^{(s)}=10^4$ nm \cite{length},
$E_c=25$ V/cm
at $T=300$ K and $E_c=0.25$ V/cm at $T=3$ K.

The physics of the field effects on the spin diffusion becomes
clearer at the strong-field limit, where
$|eE|/K_B T \gg 1/L^{(s)}$. In this limit,
the electrons move
with velocity $|E| \nu_e$ and so does the spin polarization.
$L_d$ is simply the distance over which the carriers move within
the spin life time $\tau$, $L_d  \simeq |E| \nu_e \tau =
|E| \frac{e}{k_B T} D\tau = (L^{(s)})^2 |eE|/k_B T$.
For the up-stream diffusion length $L_u$ at this limit,
$L_u \simeq k_B T /|eE|$, which simply corresponds to a Boltzmann
distribution of electrons in a retarding field.

A similar field-dependent diffusion phenomenon has been observed
and studied in
charge transport
of minority carriers in doped semiconductors \cite{book}.
In fact, if $n_{\uparrow}-n_{\downarrow}$ is substituted
by $\Delta p$ and $L^{(s)}$ is regarded as
the intrinsic
charge diffusion length, Eq. (2) becomes the diffusion equation for
the  disturbance of minority carrier  in $n$-doped semiconductors.
It is known that the electric field leads to two distinct
charge diffusion lengths in this case  as well as a
modification of carrier injection \cite{book}.

As an application of our field-dependent spin transport theory, we
study how the electric field affects
spin injection from a ferromagnetic metal to a semiconductor.
We consider a simple one-dimensional spin injection structure
to elucidate the underlying physics of
electric field and nondegenerate electron statistics effects.
This injection structure, as shown in the inset of
Fig. 3, comprises a semi-infinite metal ($x <0$) and a semi-infinite
semiconductor ($x>0$).
Electrons are injected from the metal to the semiconductor, and therefore,
the electric field is antiparallel to the $x$-axis.
In the ferromagnetic metal
the electrochemical potentials for individual spins
satisfy the equations \cite{hershfield},
$$
\frac{d^2}{dx^2}\left ( \begin{array}{c}
\mu_{\uparrow} \\
\mu_{\downarrow}
\end{array} \right )
=\left ( \begin{array}{cc}
(D^f_{\uparrow}\tau_{\uparrow\downarrow})^{-1} &
-(D^f_{\uparrow}\tau_{\uparrow\downarrow})^{-1}  \\
-(D^f_{\downarrow}\tau_{\downarrow\uparrow})^{-1} &
(D^f_{\downarrow}\tau_{\downarrow\uparrow})^{-1}
\end{array} \right )
\left ( \begin{array}{c}
\mu_{\uparrow} \\ \mu_{\downarrow}
\end{array} \right ),
$$
where $D^f_{\uparrow(\downarrow)}$ is the up-spin (down-spin) electron
diffusion constant. In metals the conductivity and the diffusion constant
are related via
$\sigma^f_{\uparrow(\downarrow)}/D^f_{\uparrow(\downarrow)}=e^2N_{\uparrow(\downarrow)}(E_F)$,
and $N_{\uparrow(\downarrow)}(E_F)$ is the up-spin (down-spin)
density of states  at Fermi energy.
It is readily seen that the above equations lead to Eq. (1) if
$L^{(f)}=[(D^f_{\uparrow}\tau_{\uparrow\downarrow})^{-1}
+(D^f_{\downarrow}\tau_{\downarrow\uparrow})^{-1}]^{-1/2}$.
The general solution can be written as
\begin{equation}
\frac{1}{eJ}\left ( \begin{array}{c} \mu_{\uparrow}\\ \mu_{\downarrow}
\end{array} \right )
=\frac{x}{\sigma^f_{\uparrow}+\sigma^f_{\downarrow}}
\left ( \begin{array}{c} 1\\ 1
\end{array} \right )+ C_1 e^{\frac{x}{L^{(f)}}} \left
( \begin{array}{c}1/\sigma^f_{\uparrow}\\ -1/\sigma^f_{\downarrow}
\end{array} \right ),
\end{equation}
where $J$ is the total electron current, which is a constant throughout the
structure
in steady state.
In the semiconductor, according to Eqs. (2) and (3),
\begin{equation}
\Delta n_{\uparrow}=-\Delta n_{\downarrow}=C_2 \exp (-x/L_d),
\end{equation}
and $J=\sigma_s E$.
In order to match boundary conditions at the interface between the metal and
the semiconductor, it is desirable to know the electrochemical potentials
for up-spin and down-spin electrons in the semiconductor, which are
related to the electron density for individual spins via
\begin{equation}
\mu_{\uparrow(\downarrow)}=k_B T \ln \Bigl(1+\frac{2
\Delta n_{\uparrow(\downarrow)}}{n_0}\Bigr) + e E x -C_0.
\end{equation}
This relation can be readily derived based on the definition of the
electrochemical potential
in nondegenerate semiconductors
$n_{\uparrow(\downarrow)}
\propto \exp [(\mu_{\uparrow(\downarrow)}
+e \psi)/k_B T]$, where $E \equiv -d \psi /dx$.

The three unknown coefficients $C_i$ ($i=0,1,2$) in Eqs. (6)-(8)
will be determined by
the boundary conditions
at the interface. For a clean and transparent interface, i.e.,
no spin-flip scattering at the interface and
no interface resistance, both the electrochemical potential and
the current for individual spins are continuous, giving rise to three
independent equations:
(1) $\mu_{\uparrow}(0^-)
=\mu_{\uparrow}(0^+)$, (2) $\mu_{\downarrow}(0^-)
=\mu_{\downarrow}(0^+)$, and (3) $j_{\uparrow}(0^-)-j_{\downarrow}(0^-)
= j_{\uparrow}(0^+)-j_{\downarrow}(0^+)$.
The current can be calculated using
$j_{\uparrow(\downarrow)}=\sigma_{\uparrow(\downarrow)}
\frac{d(\mu_{\uparrow(\downarrow)}/e)}{dx}$.

\begin{figure}
\vspace{5pt}\includegraphics[width=7cm]{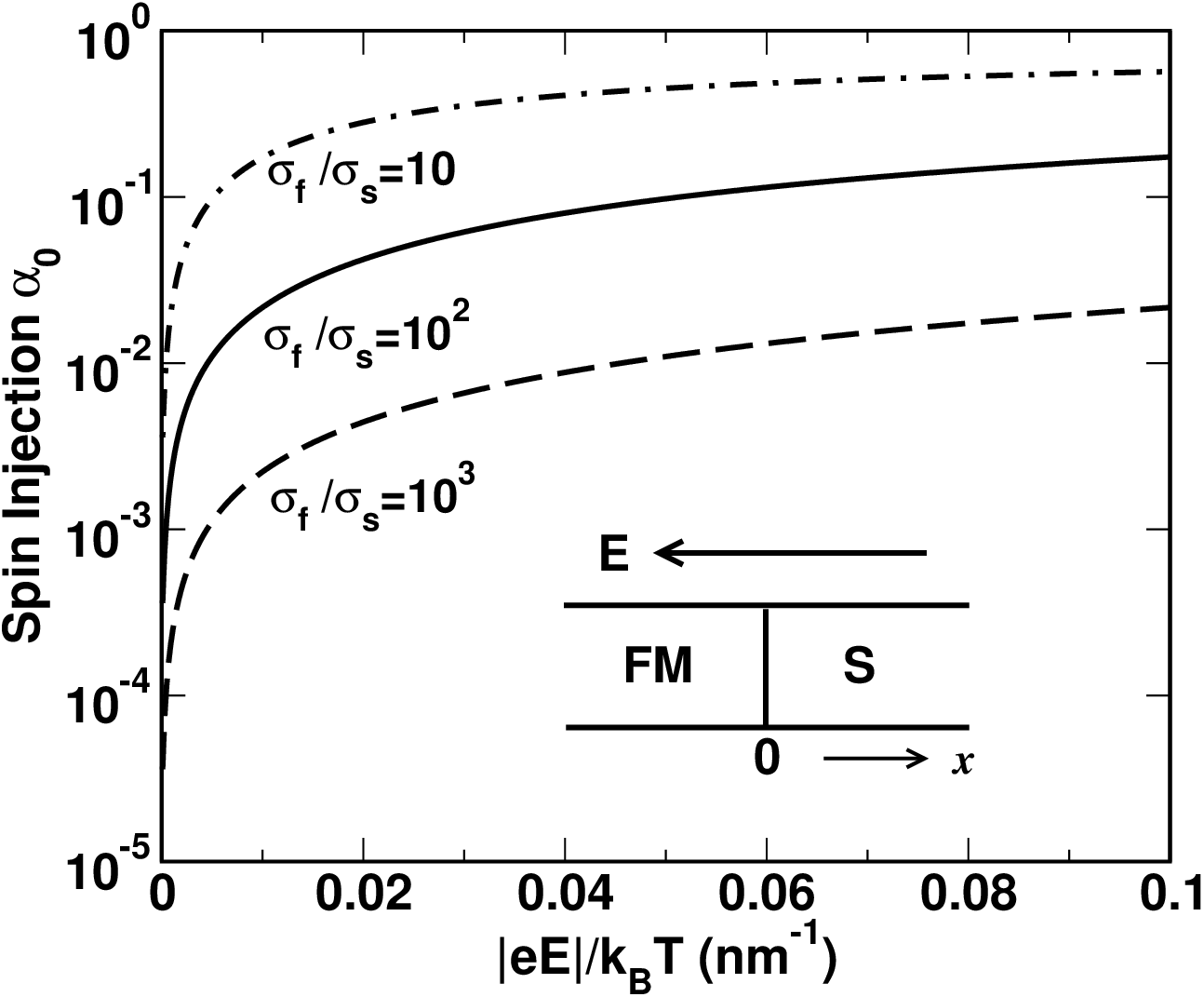}
\caption{Spin injection efficiency $\alpha_0$ as a function of electric
field.
Dot-dashed, solid, and dashed lines correspond to $\sigma_f/\sigma_s=10$,
100, and 1000, respectively. Other parameters are $p_f=0.8$, $L^{(f)}=100$ nm,
$L^{(s)}=10^4$ nm. The inset shows the schematic injection structure.}
\vspace{-10pt}
\end{figure}

The spin injection in the semiconductor is usually defined via
the spin polarization of the current,
$\alpha(x) = [j_{\uparrow}(x)-j_{\downarrow}(x)]/J$,
which is found to be proportional
to the spin polarization of
the electron density $n_{\uparrow}-n_{\downarrow}$,
\begin{equation}
\alpha (x) = \frac{n_{\uparrow}(x)-n_{\downarrow}(x)}{n_0}\Bigl(1-\frac{k_B
T}{e E L_d}\Bigr).
\end{equation}
Thus the solution of $n_{\uparrow}-n_{\downarrow}$ in Eq. (7) indicates
$\alpha (x)= \alpha_0 e^{-x/L_d}$, where $\alpha_0$ is
the spin injection efficiency.
We obtain an equation for $\alpha_0$, noting
$1-k_BT/eEL_d=-k_BT/eEL_u$,
\begin{equation}
\frac{2L^{(f)}(\alpha_0-p_f)}{(1-p^2_f)\sigma_f}=\frac{k_B T}{eE \sigma_s}
\ln \frac{-k_BT/e E L_u +\alpha_0}{-k_BT/e E L_u -\alpha_0},
\end{equation}
where  $\sigma_f=\sigma^f_{\uparrow}+\sigma^f_{\downarrow}$, and
$p_f=(\sigma^f_{\uparrow}-\sigma^f_{\downarrow})/\sigma_f$ is the spin
polarization in
the metal.
We solve Eq. (10) and plot the spin injection efficiency
$\alpha_0$ as a function of the electric field in Fig. 3.
We see that the electric field can enhance the spin injection
efficiency considerably.
When $\Delta n_{\uparrow(\downarrow)}/n_0 \ll 1$, i.e., small spin polarization
in the semiconductor,
$\alpha(x)$ can be expressed in a compact form,
\begin{equation}
\alpha(x)
=\Bigl[\frac{L^{(f)}}{(1-p^2_f)\sigma_f}
+\frac{L_u}{\sigma_s}\Bigr]^{-1} \frac{p_f L^{(f)}}{(1-p^2_f)\sigma_f}
e^{-\frac{x}{L_d}}.
\end{equation}

This remarkable expression
shows that the electric-field
effects on spin injection can be
described in terms of
the two field-induced diffusion lengths. Both diffusion lengths
affect spin injection favorably but in a different manner.
The up-stream length $L_u$ controls the relevant resistance in the
semiconductor, which
determines the spin injection efficiency. With increasing field this
effective resistance, $L_u/\sigma_s$, becomes smaller, and accordingly
the spin injection efficiency is enhanced.
The transport distance of the injected spin polarization in the semiconductor,
however, is controlled by the down-stream length $L_d$. As
the field increases, this distance becomes longer.

We now contrast Eq. (11) with that obtained by previous
calculations \cite{schmidt,rashba,smith} based on Eq. (1).
The spin injection
\begin{equation}
\alpha(x)=\Bigl[\frac{L^{(f)}}{(1-p^2_f)\sigma_f}+
\frac{L^{(s)}}{\sigma_s}\Bigr]^{-1}
\frac{p_f L^{(f)}}{(1-p^2_f)\sigma_f}e^{-\frac{x}{L^{(s)}}}
\end{equation}
is given by the zero-field result of Eq. (11).
As $L^{(f)} \ll L^{(s)}$ and $\sigma_f \gg \sigma_s$, the
effective resistance in the metal, $L^{(f)}/\sigma_f$, is much less than its
counterpart in the semiconductor, $L^{(s)}/\sigma_s$.
Thus Eq. (12) suggests that
this resistance mismatch makes
it virtually impossible to realize an appreciable spin injection
from a ferromagnetic metal to a semiconductor.
However, the more general description of the spin transport in semiconductors
indicates
that
the effective semiconductor resistance to be compared with $L^{(f)}/\sigma_f$
should be $L_u/\sigma_s$ rather than $L^{(s)}/\sigma_s$. Since $L_u$ can be
smaller than $L^{(s)}$ by orders of magnitude in the high-field regime,
this ``conductivity mismatch'' obstacle may be overcome with the help
of strong electric fields, or equivalently, large injection
currents \cite{darryl}.
For example, if the parameters of a spin injection device are as follows,
$p_f=0.8$, $L^{(f)}=100$ nm, $L^{(s)}=10^4$ nm, and $\sigma_f=100 \sigma_s$,
at zero field the spin injection efficiency is 0.04\%,
which can be increased to 4.2\% at $|eE|/k_BT=0.02$ nm$^{-1}$, which
corresponds to $|E|=50$ V/cm, or $|J|=50$ A/cm$^2$ for a typical
semiconductor conductivity $\sigma_s=1$ ($\Omega$ cm)$^{-1}$, at $T=3$ K.
This may explain the large spin injection
percentages from ZnMnSe to ZnSe \cite{molenkamp,jonker1} and
from Fe to GaAs \cite{ploog,jonker2}, as well as the
dramatic increase in spin injection with current in Ref. \cite{jonker2}.
Finally, we note that spin injection enhancement from a
 spin-selective interfacial barrier between the ferromagnetic
metal and the semiconductor, which has been identified in the low-field
regime \cite{rashba,smith}, becomes more pronounced in the high-field
regime.

In summary,
we have derived the drift-diffusion equation for spin polarization
in a semiconductor by consistently taking into account
electric-field effects and nondegenerate electron statistics.
This equation provides a framework to understand spin transport in
semiconductors.
We have identified a high-field diffusive regime which has no analogue
in metals. In this regime, there are two distinct
spin diffusion lengths, i.e., the up-stream  and down-stream
spin diffusion lengths.
The high-field description of the spin transport in semiconductors
predicts that the electric field can effectively enhance
spin injection
from a ferromagnetic metal  into a semiconductor
and substantially increase the transport distance of the spin
polarization in semiconductors.  Our results
suggest
that the ``conductivity mismatch'' obstacle in spin injection
may be overcome with the help
of high field injection in the diffusive regime.

We would like to thank N. Samarth
for pointing out the relevance of resistance mismatch for spin
injection from ferromagnetic semiconductors.
This work was supported by DARPA/ARO DAAD19-01-0490.


\begin{thebibliography}{}
\bibitem{darryl} Although it has been realized that spin injection
can be enhanced by increasing the total injection current \cite{smith},
the electric-field effects on spin transport were not taken into account and
therefore the physics of the high-field regime was not captured.
\end{thebibliography}
\end{document}